# LOW-ENERGY ELECTRON-POSITRON COLLIDER TO SEARCH AND STUDY (μ+μ−) BOUND STATE

A.V. Bogomyagkov, V.P. Druzhinin, E.B. Levichev, A.I. Milstein, S.V. Sinyatkin

*Budker Institute for Nuclear Physics, Novosibirsk 630090, Russia*

We discuss a low energy e+e− collider for production of the not yet observed (μ+μ−) bound system (dimuonium). Collider with large crossing angle for e+e− beams intersection produces dimuonium with non-zero momentum; therefore, its decay point is shifted from the beam collision area providing effective suppression of the elastic e+e− scattering background. The experimental constraints define subsequent collider specifications. We show preliminary layout of the accelerator and obtained main parameters. High luminosity in chosen beam energy range allows to study $\pi^\pm$ and $\eta$-mesons.

## 1  INTRODUCTION

Currently some authors consider a lepton atom (μ+μ−)[1] as a strong candidate for casting light on disagreement between the anomalous magnetic moment measurement and Standard Model prediction, the muonic hydrogen proton's charge radius discrepancy, etc. [2, 3]. Even if this expectation will not come true, a discovery of a new lepton bound state, which has not been observed yet contrary to positronium (e+e−) and muonium (μ+e−), and its study is by far a challenging and interesting scientific enterprise. Like positronium, dimuonium is a Bohr atom and its study (including transition spectroscopy, lifetime precise measurement, etc.) allows verifying QED and quantum mechanics computations with great accuracy. In such a research, (μ+μ−) has higher new-physics reach potential in comparison with other exotic atoms. It has larger reduced mass $\mu = m_\mu/2$ than (e+e−) and (μ+e−), for which $\mu \approx m_e$; therefore, it is considerably more compact (the dimuonium Bohr radius is 200 times smaller than the positronium one). In contrast to (pe−) and (pμ−) systems, (μ+μ−) experiences no strong interaction obscuring exploration of tiny QED effects. Study of heavier and more compact taonium (τ+τ−) would also be interesting but its production is extremely difficult even if compared with (μ+μ−).

In the past, many of proposed (μ+μ−) production channels have been discussed [1, 4-13] including those from electron and positron collision [1, 5, 10]. The e+e−→ (μ+μ−) production mechanism is particularly interesting because it contains no hadrons which, otherwise, should be untangled in the reconstruction process. However, a e+e− collider for the dimuonium experimental study has never been discussed in details.

In the paper, we propose a preliminary design of the electron-positron collider for (μ+μ−) search and study. Low beam energy $E_b \approx 400$ MeV results in a compact machine, inexpensive in both manufacture and maintenance. Large crossing angle of the intersecting beams induces a non-zero dimuonium momentum which is a distinguish feature of our design (originally proposed in

---

[1] Dubbed "dimuonium", "bimuonium" or "true muonium".

[10]). In this case, the (µ⁺µ⁻) production and decay points are spatially displaced along the beams directions bisector (for equal beam energies). The gap between the atom formation and its annihilation into e⁺e⁻ pair effectively suppresses the primary beams elastic scattering background and allows unambiguously detect the dimuonium decay.

The collider design is rather challenging, because large collision angle causes a number of effects in e⁺e⁻ beams interaction, absent in the head-on collisions, while low energy and high bunch intensity result in strong intra-beam scattering (IBS) and reduced Touschek lifetime. We discuss and estimate all these issues below.

Besides the (µ⁺µ⁻) research, the beam energy and collider configuration allow studying such processes as e⁺e⁻→ *pions* and e⁺e⁻→ η,η′ expanding substantially the proposed facility experimental program.

## 2  DIMUONIUM

Here, we briefly present the major properties of (µ⁺µ⁻) atom relevant to the collider design. More detailed information can be found elsewhere (see for instance [14] and references therein).

Dimuonium is one of the simplest hydrogen-like atoms and its spectrum in the first approximation is similar to the positronium spectrum with respectful mass substitution. However, whereas positronium cannot annihilate into e⁺e⁻, dimuonium, with its mass above the two-electron threshold, can, and this fact makes its formation and spectrum more sensitive to vacuum polarization effects caused by virtual electron-positron pairs production. Fig.1 shows dimuonium energy levels.

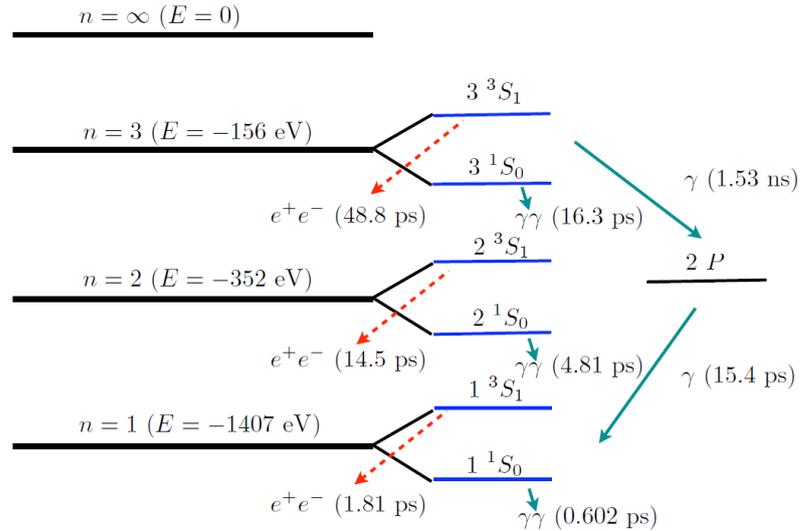

Fig.1 The dimuonium energy states diagram [10] (not in scale).

Dimuonium Bohr radius is $R_{\mu\mu}$ = 512 fm (for reference the positronium Bohr radius is $R_{ee}$ = 0.106 nm). The (µ⁺µ⁻) atom exists in two spin states: paradimuonium, which is a singlet with total muon-antimuon spin S = 0, and orthodimuonium (triplet S = 1). For paradimuonium ground state with the dominant decay mode $1^1S_0 \to \gamma\gamma$ the lifetime is 0.602 ps while for orthodimuonium it is of 1.81 ps (dominant decay is $1^3S_1 \to e^+e^-$). It is worthy of note that µ weak decay is very slow with the lifetime of 2.2 μs, and thus, dimuonium is a unique metastable system available for precision QED tests.

Cross section of (μ⁺μ⁻) production in the electron-positron collision for exact resonance condition (the beam energy $E_b = m_\mu - \Delta_{\mu\mu}/2$, $m_\mu$ is the muon mass and $\Delta_{\mu\mu} \approx 1.41$ keV is the binding energy) is [1]

$$\sigma_{\mu\mu n} = 3.3 \cdot 10^{-25} n^{-3} \text{ cm}^{-2}. \tag{1}$$

The cross section (1) rapidly decreases with the principal quantum number $n$. Radiative corrections reduce the cross section. For instance, the reduction for the ground state is [1]

$$\sigma_{\mu\mu 1R} = 0.27 \sigma_{\mu\mu 1}. \tag{2}$$

However, detailed simulation of the (μ⁺μ⁻) production rate with account for the beam energy and spatial distribution [17] recovers the cross section reduction due to the radiative corrections (within 10% accuracy), thus for convenience to estimate the dimuonium production rate below we use the cross section (1).

## 3 LARGE ANGLE BEAM CROSSING

To study the beams collision with an arbitrary angle we apply results from [15]. Two leptons with four-momenta $(E_1, \vec{p}_1)$ and $(E_2, \vec{p}_2)$ collide as shown in Fig.2. According to this definition, the head-on collision corresponds to $\theta = 0$ while for the parallel momenta $\theta = \pi/2$. The collision invariant mass is given by (we assume the speed of light $c = 1$)

$$M^2 = 2m_e^2 + 2E_1 E_2 - 2\vec{p}_1 \vec{p}_2.$$

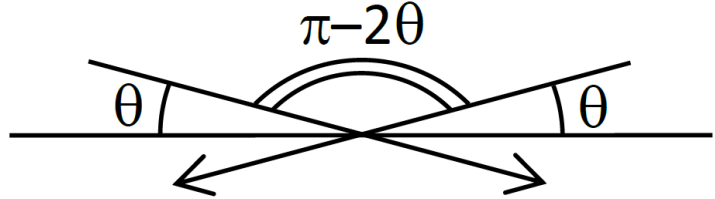

Fig.2 Electrons collision at arbitrary angle $\theta$.

In the moving reference frame conventional to circular accelerator, {x, y, z} denote the horizontal, vertical and longitudinal directions see, e.g. [16]. Introducing the normalized variables $\delta_{1,2} = (E_{1,2} - E_0)/E_0$, $x'_{1,2} = p_{x1,2}/p_0$, $y'_{1,2} = p_{y1,2}/p_0$, where $E_0$ and $p_0$ are the energy and total momentum mean values, equal for both beams, and neglecting the electron mass $m_e$, one obtains

$$M^2 = 2E_0^2 \Big[ (1 + \delta_1 + \delta_2 + \delta_1\delta_2)(1 + \cos 2\theta) + (x'_2 - x'_1 + x'_2\delta_1 - x'_1\delta_2)\sin 2\theta - y'_1 y'_2$$

$$- \frac{1}{2}(x'^2_1 + x'^2_2 + y'^2_1 + y'^2_2 - 2x'_1 x'_2)\cos 2\theta \Big]. \tag{3}$$

For the reference particles with $\delta_{1,2} = x'_{1,2} = 0$

$$M = \sqrt{2}E_0\sqrt{1+\cos 2\theta} = 2E_0\cos\theta. \tag{4}$$

Assuming Gaussian beam divergence, equation (3) allows deriving the mean and the r.m.s. values of invariant mass

$$\langle M \rangle_{x',y'} = 2E_0\left[1 + \frac{\delta_1 + \delta_2}{2} - \frac{(\delta_1 - \delta_2)^2}{8}\right]\cos\theta - \frac{E_0}{2}\sigma_{x'}^2\cos\theta - \frac{E_0}{2}\sigma_{y'}^2\frac{\cos 2\theta}{\cos\theta}, \tag{5}$$

$$\langle M^2 \rangle_{x',y'} = 2E_0^2(1 + \delta_1 + \delta_2 + \delta_1\delta_2)(1+\cos 2\theta) - 2E_0^2(\sigma_{x'}^2 + \sigma_{y'}^2)\cos 2\theta, \tag{6}$$

where $\sigma_{x',y'}$ is the r.m.s. angular spread. Equations (5) and (6) give the invariant mass energy resolution

$$\sigma_M^2 = \langle M^2 \rangle_{x',y',\delta_1,\delta_2} - \langle M \rangle_{x',y',\delta_1,\delta_2}^2 = 2E_0^2\left[(\sigma_\delta\cos\theta)^2 + (\sigma_{x'}\sin\theta)^2\right], \tag{7}$$

where $\sigma_\delta = \sigma_E/E_0$ is the relative beam energy spread. The invariant mass resolution $\sigma_M$ is essential for narrow resonance production rate $\dot{N} \propto L/\sigma_M$, where $L$ is the luminosity. Note that for large crossing angle the horizontal angular spread contribution in (7) dominates over the energy spread.

For short bunches (a *hour-glass* effect is neglected) with equal population of $N$ colliding with the frequency $f_0$ at the arbitrary angle $\theta$ the peak luminosity is

$$L_0 = \frac{N^2}{4\pi\sigma_x^*\sigma_y^*\sqrt{1+\varphi^2}}f_0, \tag{8}$$

where $\sigma_{x,y}^*$ is the transverse r.m.s. beam size at the interaction point IP (assumed to be equal for both bunches) and $\varphi = \sigma_z^*\tan\theta/\sigma_x^*$ is the Piwinski angle (here $\sigma_z^*$ is the bunch length). For large Piwinski angle $\varphi \gg 1$ the peak luminosity

$$L_0 \approx \frac{N^2}{4\pi\sigma_y^*\sigma_z^*\tan\theta}f_0 \tag{9}$$

is less than the head-on luminosity and does not depend on the horizontal beam size at the IP.

Taking into account that the dimuonium binding energy 1.41 keV is much smaller than the beam energy spread (typically $\sim 10^{-4} \div 10^{-3} E_0$), one can use a Dirac delta function for the ($\mu^+\mu^-$) production cross section

$$\sigma(x'_{1,2}, y'_{1,2}, \delta_{1,2}) = \Gamma_{\mu\mu}\sigma_{\mu\mu}\delta(M(x'_{1,2}, y'_{1,2}, \delta_{1,2}) - m_{\mu\mu}), \tag{10}$$

where $M$ is the invariant mass (3), while $m_{\mu\mu}$, $\Gamma_{\mu\mu} = 0.37\cdot 10^{-6}$ keV and $\sigma_{\mu\mu}$ are dimuonium mass, leptonic width, and the peak cross section (1), respectively. The dimuonium maximum production rate can be obtained after convolution of the cross section (10) with the differential luminosity which takes into account the angular and energy distributions of the colliding beams (details can be found in [15])

$$\dot{N}_{\mu\mu} = \frac{L_0 \Gamma_{\mu\mu} \sigma_{\mu\mu}}{2\sqrt{\pi} M \cdot \sigma_M}. \quad (11)$$

Obviously, to get the number of atoms produced over the specified time span, one needs to replace in (11) the peak luminosity (8) by the integrated luminosity.

As it was mentioned, a non-zero $e^+e^-$ crossing angle boosts the atom ($\vec{p} = \vec{p}_+ + \vec{p}_- \neq 0$) to move along the bisector of the beam lines (for equal $e^+e^-$ energies); therefore, with given dimuonium lifetime at rest, $\tau_{0\mu\mu} = \hbar/\Gamma_{\mu\mu}$, the annihilation point is separated from the formation point by

$$l = c\tau_{0\mu\mu}\beta_{\mu\mu}\gamma_{\mu\mu} = c\tau_{0\mu\mu}\tan\theta = \frac{c\hbar}{\Gamma_{\mu\mu}}\tan\theta. \quad (12)$$

## 4 BEAM-BEAM EFFECTS

For the head-on collision, the longitudinal momenta of the colliding particles do not change. For large crossing angle, the forces from the opposite bunch are not purely transverse, and depend on longitudinal coordinates of the suffering particle. For arbitrary crossing angle beam-beam effects have been considered in [18, 19, 20]. It was shown, that the transverse influence is suppressed but the longitudinal one is emphasized for large angles. The beam-beam tune shifts for flat beams ($\sigma_y \ll \sigma_x$) clearly illustrate this fact:

$$\xi_y = \frac{Nr_e}{2\pi\gamma} \frac{\beta_y^*}{\sigma_y^* \sqrt{\sigma_x^{*2} + \sigma_z^{*2}\tan^2\theta}}, \qquad \xi_x = \frac{Nr_e}{2\pi\gamma} \frac{\beta_x^*}{\left(\sigma_x^{*2} + \sigma_z^{*2}\tan^2\theta\right)}, \quad (13)$$

where $\beta_{x,y}^*$ are the betatron functions at the IP and $r_e$ is the electron classical radius, and

$$\xi_z = -\frac{Nr_e}{2\pi\gamma} \frac{\sigma_{z0}^* \tan^2\theta}{\sigma_{\delta 0}\left(\sigma_x^{*2} + \sigma_z^{*2}\tan^2\theta\right)} \frac{\alpha}{|\alpha|}, \quad (14)$$

where $\sigma_{\delta 0}$ and $\sigma_{z0}$ are the energy spread and the bunch length of the "strong" beam affecting the test particle and keeping constant (weak-strong approximation). The bunch population $N$, relativistic factor $\gamma$, and the IP beam sizes $\sigma_x^*$ and $\sigma_z^*$ also relate to the "strong" bunch. Factor $\alpha/|\alpha|$ defines the sign of the momentum compaction.

For some (not necessary large) intersection angle $\theta$ we have $\sigma_z^* \tan\theta \gg \sigma_x^*$, and the conventional transverse beam-beam effects (13) are suppressed. Longitudinal beam-beam effects (14) increase, the synchrotron tune shift

$$\Delta v_s = \xi_z \approx -\frac{Nr_e}{2\pi\gamma} \frac{1}{\sigma_{\delta 0}\sigma_z^*} \frac{\alpha}{|\alpha|}, \quad (15)$$

where we assume $\sigma_{z0} = \sigma_z$, might be as large as the unperturbed synchrotron tune $\nu_s$, causing the phase instability of the test bunch. Even below this limit, when the particle motion is still stable, the "strong" bunch electric field distorts the accelerating RF bucket, modifies the longitudinal bunch distribution, and reduces the luminosity.

In canonical coordinates $(z, p_z)$, where z is the particle position with respect to the bunch center and $p_z$ is the conjugate momentum, the linear Hamiltonian of the test particle longitudinal motion has the form (see Attachment 1)

$$H_z = -\frac{\alpha p_z^2}{2} + \langle U_{rf} \rangle + \langle U_{bb} \rangle \approx -\frac{\alpha p_z^2}{2} - \frac{\nu_s^2}{\alpha R^2}\frac{z^2}{2} - \frac{2\xi_z \nu_s}{\alpha R^2}\frac{z^2}{2}. \tag{16}$$

Here the second term relates to accelerating RF system, the third one corresponds to the colliding bunch electrical field, α is momentum compaction factor, and *R* is the orbit average radius. To keep the bunch field effect negligible one needs $\xi_z \ll \nu_s/2$, which means either large synchrotron tune $\nu_s$ or small longitudinal beam-beam parameter (15) and finally limits the "strong" bunch population according to

$$N < \frac{2\pi R \cdot \gamma}{r_e} \frac{\alpha \cdot \sigma_\delta^2}{2}. \tag{17}$$

This limit within the factor of two corresponds to the condition for coherent instability of two bunches intersecting at large angle found in [18].

From the Hamiltonian (16) one can see another possibility to suppress the longitudinal beam-beam effects. For positive momentum compaction $\alpha > 0$ the beam-beam synchrotron tune shift $\xi_z < 0$, and the "strong" bunch field flattens the RF potential well, increases the "weak" bunch length, and destroys stability. Alternatively, for magnetic lattice with $\alpha < 0$ the third term in (16) has the same sign as a second one, giving deeper RF potential well and smaller "weak" bunch length.

High order expansion of the synchrotron motion Hamiltonian shows that for $\alpha > 0$ the "strong" bunch field modulates the RF potential well and splits the "weak" bunch to many sub-bunches (see Attachment 1). Simulation with the help of the beam-beam computer code LIFETRAC [23] confirmed this effect as well as above estimation of the bunch intensity limit.

## 5  COLLIDER PARAMETERS SELECTION

In general, decrease of the collider energy results in simplification of the systems (magnetic, accelerating RF, vacuum, etc.), size and cost reduction, shortening of implementation period, maintenance cost reduction, etc. Our first proposal of the e⁺e⁻ collider for (μ⁺μ⁻) production was based on the head-on collision design with the beam energies of 105 MeV [21]. The double ring facility, called μμ-*tron* had the overall size ≈ 5×10 m², the peak luminosity of ~$10^{30}$ cm⁻²s⁻¹, 10 keV invariant mass resolution with collision monochromatization à la A. Renieri [22], provided the dimuonium maximum production rate of ~50 atoms per hour. We had intended to observe the dimuonium by searching for X-rays from (μ⁺μ⁻) Bohr transitions such as 2P→1S [5]. However, production and rapid annihilation of a single neutral system at rest inside the bunch are difficult to identify and separate from the dominating QED background, and we failed to find the relevant detecting system. Also, a weak point of very low collider energy is low radiation damping, strong intra-beam scattering (IBS), and beam lifetime degradation due to the Touschek effect. Furthermore, low energy collision is inefficient from a viewpoint of the injection facility: in order to get sufficient positron rate after converter, the primary electron beam requires much higher energy than that in the collider.

Moderate increase of the collider energy overcomes (or at least mitigates) these problems: the radiation damping intensifies, IBS increment reduces, Touschek life time increases. Also, it gives possibility of the top-up injection at the energy of the experiment. The e⁺ and e⁻ beams can be merged at appropriate angle to boost the (μ⁺μ⁻) atoms and provide their decay to leptons far from the primary beams collision area effectively decreasing elastic scattering background e⁺e⁻→e⁺e⁻.

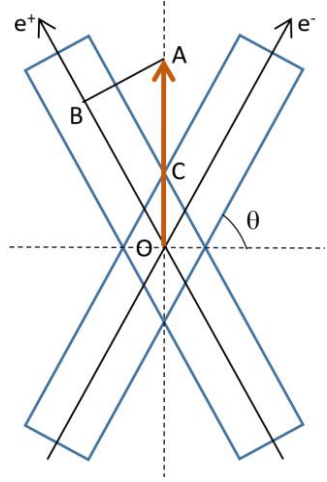

Fig.3 On estimation of the background rate.

To estimate the signal-background ratio $\dot{N}_{\mu\mu}/\dot{N}_{ee}$, where $\dot{N}_{ee}$ is the elastic scattering rate, we use the collision scheme shown in Fig.3. Here points O and A are those of dimuonium production and decay; the distance $OA = l$ is defined in (12). According to (11) the (μ⁺μ⁻) production rate is

$$\dot{N}_{\mu\mu} \propto N_0^2 / \sigma_M .$$

In the point A, where e⁺e⁻ elastic scattering mimics the dimuonium decay to leptons, the density of the primary particles (normal distribution is assumed) falls off according to

$$N_1 = N_0 \exp\!\left[-n_x^2/2\right],$$

where $n_x = AB/\sigma_x$, $AB = l \cdot \cos\theta = c\tau_{0\mu\mu} \sin\theta$, and the background rate is $\dot{N}_{ee} \propto N_1^2$.

Using $\sigma_M$ from (7) and expressing the beam size and divergence through the emittance and beta function $\sigma_{x'}^2 = \varepsilon_x / \beta_x$ and $\sigma_x^2 = \varepsilon_x \beta_x$ we obtain the following estimation

$$\frac{\dot{N}_{\mu\mu}}{\dot{N}_{ee}} \propto \frac{\exp\left[\frac{(c\tau_{0\mu\mu})^2 \sin^2\theta}{\sigma_x^{*2}}\right]}{\sqrt{\sigma_\delta^2 \cos^2\theta + (\sigma_x^{*2}/\beta_x^{*2})\sin^2\theta}}, \quad (18)$$

This estimate allows defining collider requirements. To increase numerator in (18) one needs the large crossing angle $\sin\theta \to 1$ ($\theta \to 90°$) and small horizontal beam size $\sigma_x^2 = \varepsilon_x \beta_x$ at the IP. To reduce denominator (better resolution) one needs small beam divergence $\sigma_{x'}^2 = \varepsilon_x/\beta_x$ at the IP. Large intersection angle increases required beam energy (4), enlarges the energy resolution (7), and strongly degrades the peak luminosity (9). These disadvantages are intrinsic for the Fool's Intersection Storage Ring in [10]: for $\theta = 85°$ and beam energy $E_0 \approx 1.2$ GeV, the luminosity would be low, the energy resolution would be large and the size of the facility could hardly be compact.

Here we specify μμ-tron energy, beam crossing angle, orbit length, and other parameters balancing pros and cons considered above. An important issue in our case is existing injection facility [24] with capacity ~0.5÷1·10$^{10}$ e$^+$/s in the energy range of $E_{inj} \approx 350 \div 450$ MeV. Top-up injection (at the energy of experiment) allows reaching average luminosity close to the peak one. Despite the luminosity drop with the crossing angle increase, reduction of the beam interaction length allows decreasing of the vertical beta $\beta^*_y$ at the IP (like for Crab-Waist collision [25]), this partially recovers the luminosity loss. In addition, the beam energy enlargement suppresses IBS and results in the collider performance enhancement.

The major specifications of the electron-positron collider for dimuonium production and exploration are:

1. The crossing angle of 75° allows ($\mu^+\mu^-$) production for the beam energy of $E_b = 408$ MeV ($m_\mu = 105.7$ MeV). At 70° $\pi^+\pi^-$ ($m_{\pi\pm} = 139.6$ МэВ) can be generated at the same beam energy. The angle modification is available either by mechanical rearrangement of the interaction area or using corrector magnets.
2. The crossing angle of 75° provides $1^3S_1$ ($\mu^+\mu^-$) decay path $l = 2$ mm, it is enough to detect the atom decay to e$^+$e$^-$. For the next dimuonium terms, the lifetime (and the escape length) increases as $n^3$ (see Fig.1), however production rate falls down insofar.
3. To increase the signal-background ratio (18) the horizontal beam size at the IP should be $\sigma_x^* << c\tau_{o\mu\mu} = 0.54$ mm, so we take $\sigma_x^* \leq 0.15$ mm.
4. For large crossing angle, horizontal angular spread at the IP is the main source of the invariant mass resolution degradation. To reduce the angular spread having the small beam size collider has to provide low horizontal emittance.
5. Lattice with negative momentum compaction α<0 have been studied at several colliders (for instance, at КЕК В [26] or at ДАФNЕ [27]). This operation mode has never been applied as the regular one. The explanation is that in such a mode microwave instability current threshold reduces, giving larger beam energy spread and transverse size (especially in vertical direction) and smaller luminosity [27]. Therefore, in our design we adopt a conservative but robust solution with positive α. Another argument in favor of α>0 is that the negative compaction lattice requires extra space for matching sections and the orbit length increases.

6. For high luminosity one needs highly populated $e^+e^-$ bunches but the longitudinal beam-beam effects limit the bunch intensity according to (17). In our specifications we take the bunch population $N_0 \leq 3.5 \cdot 10^{10}$ $e^+/e^-$. To arrange a multi-bunch operation two separate storage rings are necessary.

7. Low vertical beta-function at the IP benefits to the high luminosity. However, extremely low beta causes strong chromatic effect in the final focus quadrupoles, which needs to be compensated by specially arranged chromatic correction sections. To reduce collider size we choose $\beta^*_y = 2$ mm and avoid final focus local chromaticity correction.

8. Reverse of the beam circulation direction in one of the rings changes the crossing angle to 15°, and makes it possible to study with high luminosity energy range from η-meson ($m_\eta$ = 547.8 MeV, $E_b \approx$ 284 MeV at θ = 15°) to η′-meson ($m_{\eta'}$ = 957.7 MeV, $E_b \approx$ 496 MeV).

## 6     COLLIDER

A simple and symmetric configuration of the large crossing angle collider consists of two identical bulb-shape storage rings with two intersection points. The $e^+e^-$ orbits for such configuration are shown schematically in Fig.4. Each ring includes two straight interaction sections and two closing arcs: inward and outward. The collider footprint is ≈12×6 м², and the ring circumference is about 23 m.

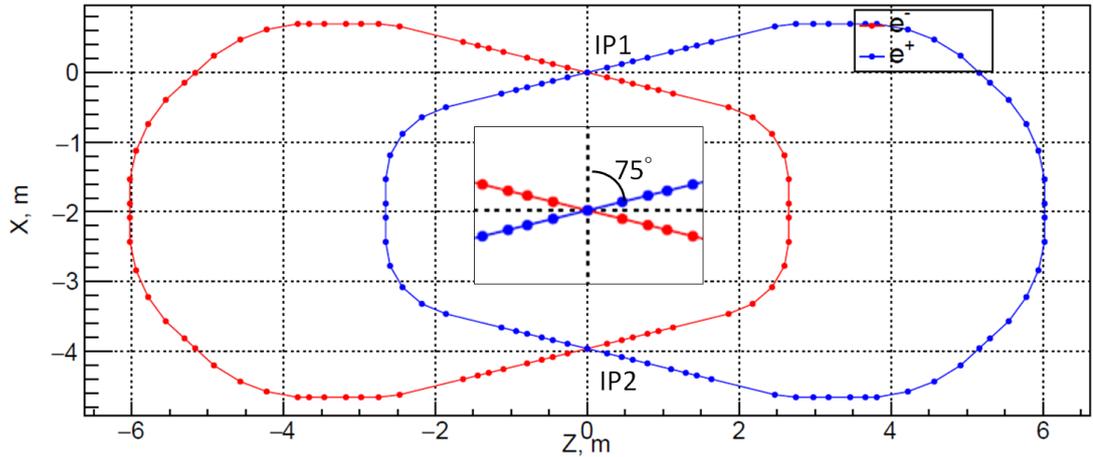

Fig.4 Orbits of $e^+e^-$ collider rings.

Fig.5 shows the magnet sequence and the lattice functions along the ring. The inward arc lattice (IP1-IP2) is a Double Bend Achromat (DBA) with the IP betatron functions $\beta^*_y = 2$ mm and $\beta^*_x = 20$ cm. The maximum vertical beta in the first defocusing final focus quadrupole is not large ($\beta_y \approx 50$ m) and the local chromatic section is unnecessary. The outward Multipole Bend Achromat (MBA) arc (IP2-IP1) provides low radiative emittance [28]. Table 1 gives the main parameters of the ring at 408 MeV.

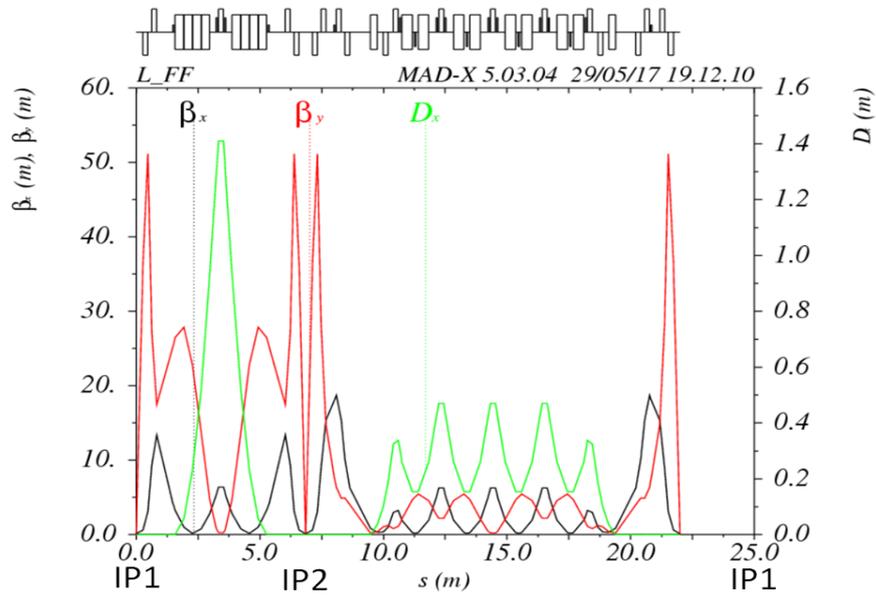

Fig.5 Storage ring lattice functions.

Table 1 μμ-tron main parameters at 408 GeV

| Beam energy (MeV) | 408 |
|---|---|
| Circumference (m) | 23 |
| Bunch intensity/current (mA) | $3.5\times10^{10}$/73 |
| Revolution frequency/period (MHz)/(ns) | 13.04/76.7 |
| RF harmonic number/frequency (MHz) | 26/338.98 |
| Energy loss per turn (keV) | 2.3 |
| RF voltage (kV) | 450 |
| RF acceptance | 2% |
| Synchrotron tune | $1.71\times10^{-2}$ |
| Momentum compaction α | $6.4\times10^{-2}$ |
| Damping time hor/ver/long (ms) | 17.3/27.3/22.1 |
| Damping partition hor/long | 1.6/1.4 |
| Horizontal emittance (without/with IBS) (nm) | 26/90 |
| Energy spread (without/with IBS), $\times10^4$ | 4/8.4 |
| Bunch length (without/with IBS) (mm) | 5.4/11.6 |
| Betatron coupling | 0.3% |
| IP horizontal angular spread $\sigma^*_{x'}\times10^4$ | 6.7 |
| Invariant mass resolution (keV) | 390 |
| Hor/vert betatron function at IP (mm) | 200/2 |
| Hor/vert betatron size at IP (μm) | 130/0.7 |
| Hor/vert beam-beam parameter ($\xi_x/\xi_y$) | $2\times10^{-6}/1.2\times10^{-3}$ |
| Longitudinal beam-beam parameter $\xi_z$ | $-2\times10^{-3}$ |
| Peak luminosity for 1 bunch (cm$^{-2}$s$^{-1}$) | $4\times10^{30}$ |
| Peak luminosity for 20 bunches (cm$^{-2}$s$^{-1}$) | $8\times10^{31}$ |

The momentum compaction factor is quite large α = 0.064, so that condition $\xi_z \ll \nu_s/2$ is well satisfied. The longitudinal beam-beam effects define the maximum bunch intensity (17) as $N_{max} = 1.5 \times 10^{11}$. We have chosen $N = 3.5 \times 10^{10}$ and it seems rather adequate for reliable operation. The transverse beam-beam effects are negligible and allow choice of the betatron tune point from other reasons (for instance, size of dynamic aperture or momentum acceptance). The lattice provides radiative emittance of $\varepsilon_x$ = 26 nm which is increased by IBS by factor of four. IBS elongates the bunch length from natural 5.4 mm to 11.6 mm. Touschek beam lifetime for the parameters given in Table 1 is around 1500 s.

The primary e⁺e⁻ collisions beyond the point A in Fig.3 determine elastic scattering background rate. Assuming the Gaussian distribution of the beam particles along the dimuonium escape line with standard deviation of $OC = \sigma_x^*/\cos\theta$, the number of the electrons in the beam tail beyond the A divided by the total amount of beam electrons is

$$N_{>A}/N_\Sigma \propto 1 - erf\left(\frac{OA}{\sqrt{2}\cdot OC}\right) = 1 - erf\left(\frac{l\cos\theta}{\sqrt{2}\sigma_x^*}\right).$$

Our parameters provide OA>2 mm and suppress the particle density in the beam tail (and the elastic scattering background rate) by 4.5 standard deviations for the ground state $1^3S_1$.

There are eight bending magnets in each ring. Two magnets compose the DBA cell of the inward arc. Six magnets compose the regular MBA cell in the outward arc and two additional magnets cancel dispersion function in the interaction region. All bending magnets have the same magnetic field (1.27 T at 408 MeV) and low negative gradient to reduce the number of defocusing quadrupoles, to increase horizontal partition number and to reach smaller emittance. The regular magnets in the MBA cell are split into two halves with defocusing sextupoles inserted in between, at the azimuth with the best ratio of the vertical-to-horizontal betas. Twenty quadrupoles combined in 10 families determine the ring optical functions. The maximum gradient in the arc quadrupoles is rather tolerable (≈15 T/m) but the first (defocusing) final focus quadrupole QD0, providing the vertical beta $\beta_y^*$ = 2 mm at the IP, is challenging: it should combine the maximum gradient of –32 T/m with extreme compactness to accommodate the detector equipment.

Fig. 6 shows an artist sketch of the interaction area. It consists of the experimental vacuum vessel with the beams intersecting inside, final focus quadrupoles, vacuum tubes and equipment (pumps, valves, bellows, etc.), beam diagnostics equipment (beam position monitors). The experimental chamber is a flat box with 0.5-mm-thick beryllium windows on the box top and bottom to pass e± produced by the dimuonium atoms decay. Presently detector set up for (μ⁺μ⁻) study looks simple and compact; it consists of the tracking systems above and below the median plane, magnetic spectrometer, etc. Four thin-walled stainless steel vacuum tubes with QD0 defocusing quadrupoles over them join the experimental chamber. The quadrupoles are built of permanent magnets and provide the maximum gradient up to –35 T/m in the ⌀30 mm aperture while being very compact and allowing enough room for passing μ⁺μ⁻ (or π⁺π⁻) from the IP toward the magnetic spectrometer. The next quadrupole (denoted in Fig.6 as QD/QF1) is a conventional electromagnet and serves either as focusing (at the low energy) or as additional defocusing (at the high energy) quadrupole.

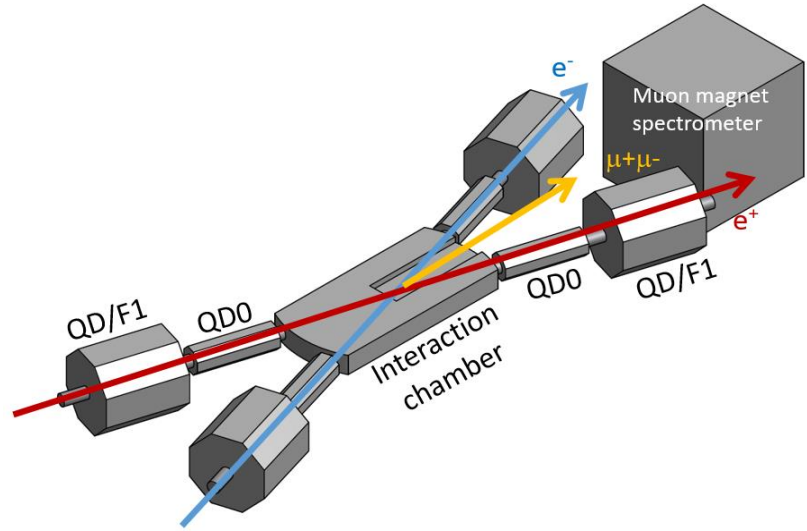

Fig.6 Schematic arrangement of the interaction region equipment.

Table 2 clearly demonstrates that production rate of dimuonium allows observation of the atom in the reasonable time and investigation of its properties, including spectrum. Fig.7 illustrates ($\mu^+\mu^-$) states distribution along the escape line.

Table 2 Dimuonium production rate estimation for collider parameters from Table 1

| ($\mu^+\mu^-$) rate | 1 hour | 4 months |
|---|---|---|
| Totally (1S/2S/3S) | 65/8/2.4 | 124000/16000/4600 |
| For $l$ >2 mm | 19/5/1.6 | 55000/14000/4500 |

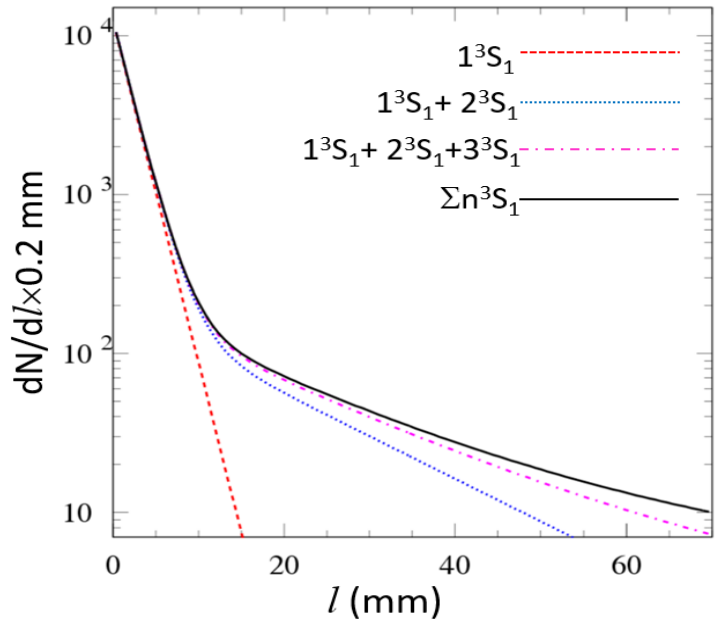

Fig.7 ($\mu^+\mu^-$) distribution along the escape line.

As it was mentioned above, the beam energy 408 MeV under the crossing angle of 70° allows formation of π⁺π⁻ pairs. The collider parameters are the same as for 75°. To study pions two scenarios are possible: (a) both collision points are modified mechanically from 75° to 70° after the dimuonium research program is over, (b) simultaneous operation when the beams collide at 75° in the first point to study (μ⁺μ⁻) and at 70° in the second point to study π⁺π⁻.

## 7   REVERSE BEAM CONFIGURATION

Reverse of one of the beams changes collision angle from 75° to 15° and opens additional possibilities for experiments. In the injection facility energy range (~300÷500 MeV) collider with 15° beam crossing allows experiments in the range of $E_{CM} \approx 500 \div 1000$ MeV which includes ρ, ω, η, and η′-mesons, etc. Such experiments require a general-purpose detector with solenoid field, which is very much different from the (μ⁺μ⁻) detector set up. Thus, the interaction region needs reconfiguration including vacuum chamber, final focus magnets, etc. toward the conventional e⁺e⁻ colliders arrangement. In addition, injection polarity in one of the rings should be reversed. As there are no crucial difficulties for that, we do not consider the new configuration in detail and for further estimation apply the lattice from the previous section. In what follows, we assume adjustment of the new lattice with 15° crossing angle to the optics in Fig.5 and estimate collider parameters by scaling the energy (obviously, the IBS is taken into account). Table 3 shows new collider specifications (excluding those repeating the Table 1: circumference, beam current, RF frequency, etc.) for the minimum (η-meson, $m_\eta$ = 547.8 MeV, $E_b \approx 284$ MeV) and maximum (η′-meson, $m_{\eta'}$ = 957.7 MeV, $E_b \approx 496$ MeV) beam energies.

Table 3 Main collider parameters for 15° collision angle.

| Beam energy (MeV) | 283.59 (η) | 495.78 (η′) |
|---|---|---|
| Invariant mass (MeV) | 547.86 | 957.76 |
| Energy loss per turn (keV) | 0.535 | 4.997 |
| RF voltage (kV) | 300 | 550 |
| Synchrotron tune | $1.67 \times 10^{-2}$ | $1.71 \times 10^{-2}$ |
| Horizontal emittance (without/with IBS) (nm) | 11.4/105 | 34.8/75 |
| Energy spread (without/with IBS), $\times 10^4$ | 2.8/10.6 | 4.8/8.4 |
| Bunch length (without/with IBS) (mm) | 3.7/14.2 | 6.3/11 |
| IP horizontal angular spread $\sigma^*_{x'} \times 10^4$ | 8.3 | 7.1 |
| Invariant mass resolution (keV) | 420 | 580 |
| Hor/vert beam-beam parameter ($\xi_x/\xi_y$) | $3 \times 10^{-4}/1.4 \times 10^{-2}$ | $3 \times 10^{-4}/1.3 \times 10^{-2}$ |
| Longitudinal beam-beam parameter $\xi_z$ | $-1.8 \times 10^{-3}$ | $-1.7 \times 10^{-3}$ |
| Peak luminosity for 1 bunch (cm⁻²s⁻¹) | $3.3 \times 10^{31}$ | $5.2 \times 10^{31}$ |
| Peak luminosity for 20 bunches (cm⁻²s⁻¹) | $6.6 \times 10^{32}$ | $1 \times 10^{33}$ |

According to (9) the acute intersection angle significantly increases the luminosity. The transverse beam-beam effects although increased are still rather low compare to the head-on collision. The change of longitudinal beam-beam effects is small relatively to the case of 75° collision.

# 8 CONCLUSION

We have considered preliminary design of the electron-positron collider for production and study of the ($\mu^+\mu^-$) bound state and capable to perform other experiments with high luminosity in the center-of-mass energy range $E_{CM} \approx 500 \div 1000$ MeV. Two rings collider with large crossing angle is quite compact, the ring orbit length ~$20 \div 30$ m, and not expensive in development and maintenance. Despite many challenging issues still needing investigation (dynamic aperture, collective effects, magnet and vacuum systems design, etc.); the project seems interesting for both fundamental research and collider technologies development.

Collider performance and dimuonium production rate allow not only detection of new leptonic atoms for the first time but also study ($\mu^+\mu^-$) properties (including precise measurement of the lifetime, spectroscopic experiments, atoms passing through a thin foil, etc.).

# 9 ACKNOWLEDGEMENTS

Authors are grateful to

- Yu.A. Pupkov for his important advice to consider extremely low energy $e^+e^-$ collider,
- N.A. Vinokurov, S.I. Seredniakov, E.P.Solodov for many fruitful discussions and valuable comments,
- E.A. Perevedentsev for introduction to and D.N. Shatilov for simulation of the longitudinal beam-beam effects,
- V.A. Kiselev for estimation of the injection schemes and efficiency,
- P.D. Vobly for preliminary design of the quadrupole with permanent magnets.

# LONGITUDINAL EFFECTS IN COLLIDING BEAMS WITH ARBITRARY CROSSING ANGLE

In canonical variables $(z, p_z)$

$$z = s - ct, \qquad p_z = \frac{\Delta E}{p_0 c},$$

a synchrotron motion Hamiltonian averaged over the fast betatron oscillation has the form

$$H = -\alpha \frac{p_z^2}{2} + \langle U_{rf} \rangle + \langle U_{bb} \rangle_{x,y=0}, \qquad (A.1)$$

where the first term relates to the "kinetic" energy while the second and third terms describe the accelerating RF system and the colliding beam field respectively. For the test particle crossing the counter bunch at the angle $\theta$, the bunch potential is written as [29]

$$U_{bb} = -\frac{Nr_e(1+\cos 2\theta)}{2\pi^2} \iiint \frac{du_x du_y du_z}{u_x^2 + u_y^2 + u_z^2} \exp\left[-\frac{u_x^2 \sigma_x^2}{2} - \frac{u_y^2 \sigma_y^2}{2} - \frac{\gamma^2 u_z^2 \sigma_z^2}{2}\right.$$
$$\left. + iu_x(x\cos 2\theta + s\sin 2\theta) + iu_y y + iu_z \gamma(s(1+\cos 2\theta) - x\sin 2\theta - z)\right], \qquad (A.2)$$

where $r_e$ is the classical electron radius, $N$ is the particle number in the bunch and $\sigma_{x,y,z}$ are the r.m.s bunch sizes at the IP. Contribution of the RF system is described by

$$U_{rf} = \frac{U_0}{p_0 c \cdot \Pi}\left[z + \delta(s-s_0)\frac{\lambda_{rf}}{\sin\Phi_s}\cos\left(\Phi_s + 2\pi\frac{z}{\lambda_{rf}}\right)\right], \qquad (A.3)$$

where $U_0$ is the particle energy loss per turn, $\lambda_{rf}$ and $\Phi_s$ are the RF wavelength and equilibrium phase, $\Pi$ is the reference orbit length. Inserting (A.3) and (A.2) in (A.1) and expanding the result as a power series in $z$ up to the second order (small synchrotron oscillation approximation), after some algebraic manipulations, one gets

$$H = -\alpha \frac{p_z^2}{2} - \frac{z^2}{2}\left(\frac{k_s^2}{\alpha} - \frac{2Nr_e}{\gamma \Pi \sigma_z^2}\frac{\varphi^2}{1+\varphi^2}\right), \qquad (A.4)$$

where $\varphi$ is the Piwinski angle introduced in Chapter 3 and the synchrotron oscillation wave number is defined as

$$k_s^2 = \left(\frac{\omega_s}{R}\right)^2 = \frac{2\pi\alpha}{\lambda_{rf}\Pi}\frac{U_0}{p_0 c}\cot\Phi_s,$$

где $R$ is the average orbit radius and $\nu_s$ is the synchrotron tune.

According to (A.4), the longitudinal potential well is parabolic near the stable fix point and colliding beam field either flattens or deepens it dependently on the sign of the momentum compaction α, therefore the bunch either stretches or contracts. Higher terms in the expansion (A.4) will split the beams into sub-bunches, degrade the luminosity or, for large current, produce instability of the opposed bunch. This effect is similar, to some extent, to the third harmonic RF cavity installed in storage rings to elongate the bunches [30]. Computer simulation confirms analytic estimations of this section well.

Introducing the corrected wave number $k_s + \Delta k_s$ for longitudinal motion disturbed by the colliding bunch field, from (A.4) one can find the synchrotron motion tune shift as

$$\xi_z = \Delta \nu_s = \Delta k_s R = -\frac{Nr_e}{2\pi\gamma}\frac{\varphi^2}{1+\varphi^2}\frac{\alpha}{k_s\sigma_z^2}. \quad (A.5)$$

It is worth to note that

$$\frac{\alpha}{k_s} = \frac{\alpha}{|\alpha|}\frac{\sigma_{z0}}{\sigma_{\delta 0}},$$

where lower index «0» denotes unperturbed values.

Assuming negligible bunch elongation ($\Delta \nu_s \ll \nu_s / 2$) one can get a condition for the particle number in the colliding beam from (A.5):

$$N \ll \frac{\gamma\Pi}{r_e}\frac{k_s^2}{\alpha}\frac{\sigma_z^2}{2}\frac{1+\varphi^2}{\varphi^2}. \quad (A.6)$$